\documentclass[runningheads]{llncs}
\usepackage[T1]{fontenc}
\usepackage{graphicx}

\usepackage{soul}
\usepackage{url}
\usepackage[hidelinks]{hyperref}
\usepackage[utf8]{inputenc}
\usepackage[small]{caption}

\usepackage{amsmath}
\usepackage{amsthm}
\usepackage{booktabs}
\usepackage{algorithm}
\usepackage{algorithmic}
\usepackage[switch]{lineno}
\usepackage{hyperref}
\usepackage{enumerate}
\usepackage{graphicx}
\usepackage{enumitem}
\usepackage{amssymb}
\usepackage{multirow}
\usepackage{tabularx}
\usepackage{subcaption}
\usepackage{wrapfig}
\usepackage{bm}
\usepackage{xcolor}
\usepackage{caption}
\DeclareCaptionFont{blue}{\color{blue}} % 定义蓝色字体

\begin{document}
\title{DSparsE: Dynamic Sparse Embedding for Knowledge Graph Completion}
\author{Chuhong Yang\inst{1} \and
Bin Li\inst{1} \and
Nan Wu\inst{1}}
\authorrunning{C. Yang et al.}
\institute{Beijing Institute of Technology, Beijing, China \inst{1} \\
\email{\{3120230733, binli, wunan\}@bit.edu.cn}\\
% \url{http://www.springer.com/gp/computer-science/lncs} 
}
\maketitle              % typeset the header of the contribution
\begin{abstract}
    Addressing the incompleteness problem in knowledge graph remains a significant challenge. Current knowledge graph completion methods have their limitations. For example, ComDensE is prone to overfitting and suffers from the degradation with the increase of network depth while InteractE has the limitations in feature interaction and interpretability.
    To this end, we propose a new method called dynamic sparse embedding (DSparsE) for knowledge graph completion.
    The proposed model embeds the input entity-relation pairs by a shallow encoder composed of a dynamic layer and a relation-aware layer.
 Subsequently, the concatenated output of the dynamic layer and relation-aware layer is passed through a projection layer and a deep decoder with residual connection structure. 
 This model ensures the network robustness and maintains the capability of feature extraction.
    Furthermore, the conventional dense layers are replaced by randomly initialized sparse connection layers in the proposed method, which can mitigate the model overfitting.
  Finally,
 comprehensive experiments are conducted on the datasets of FB15k-237, WN18RR and YAGO3-10.
 It was demonstrated that the proposed method achieves the state-of-the-art performance in terms of Hits@1  compared to the existing baseline approaches.
 An ablation study is performed to examine the effects of the dynamic layer and relation-aware layer, where the combined model achieves the best performance.

\keywords{Knowledge Graph  \and Graph Completion \and Link Prediction \and Sparse Embedding.}
\end{abstract}
\section{Introduction}
Knowledge Graph (KG) is a directed heterogeneous graph that represents concepts, entities, and their relationships in a structured form using knowledge triples. Knowledge triples are typically represented as $(s, r, o)$, where $s, r$ and $o$ denote the subject entity, the relation, and the object entity, respectively. KGs have a wide range of applications in various fields, including natural language processing, information retrieval, recommendation systems, and semantic web technologies. They are used to represent and organize knowledge in a structured and machine-readable format, which can be used to power intelligent applications and services.

Some well-known KGs, including Wikidata~\cite{2014Wikidata} and DBpedia~\cite{S2007DBpedia}, contain billions of knowledge triples, but they are often incomplete, which poses a significant challenge in the field of knowledge graph research. To address this issue, knowledge graph completion has emerged as an important task, which aims to predict missing knowledge triples. Typically,  link prediction that focuses on predicting the missing entity in a knowledge triple is adopted  for knowledge graph completion. Graph embedding, which uses low-dimensional, dense, and continuous vectors to represent nodes and relationships in knowledge graphs, is the basis of most link prediction methods.
Existing link prediction models can be categorized into tensor decomposition models~\cite{kolda2009tensor}, translational models, and deep learning models~\cite{rossi2021knowledge}. Recently, pre-trained language models, such as Large Language Model, have also been introduced to solve KG incompletion problem \cite{he2023mocosa,zhu2021neural}.

 In this paper, we propose a deep learning model, called  DSparsE, for KG completion, where a new model that includes a relation-aware layer and a dynamic layer to extract features and  residual connections used for decoding is proposed for link prediction. The previous multilayer perceptron (MLP) model ComDensE \cite{kim2022comdense} retains both shared fully connected layers and relation-aware fully connected layers, and concatenates their results by a projection layer to achieve feature fusion. The relation-aware layer can be seen as a MLP with dynamic weights that changes with the input data. However, this dynamic processing is not comprehensive because the weights of the shared layers are still fixed, which limits the network's expressive power. MoE ~\cite{shazeer2017outrageouslyMoE,jacobs1991adaptive} and CondConv ~\cite{yang2019condconv} were proposed in 2017 and 2019, respectively. The former divides the fully connected layer into several expert layers and uses a separate network to generate the combination weights of these expert layers. It takes the expert blocks with the top $k$ weights for feature fusion. The latter uses dynamic convolution kernels based on input data for convolution operations. These dynamic methods give the network greater flexibility and have been shown to have good application potential.
Thus DSparsE introduces a dynamic structure similar to MoE into the encoding end, and takes the results of all expert blocks for weighted fusion.
%  This improves performance without a significant increase of parameters.

 Compared to fully connected networks, convolutional layers introduce position-related sparse connections, which suppress overfitting effectively, save computing resources, and  capture feature correlations between adjacent pixels efficiently. However, in link prediction for knowledge graph completion, the input of neural network is a one-dimensional embedding vector, which does not naturally have correlation information like pixels in images. Most of the aforementioned convolution-based models ~\cite{dettmers2018convolutionalConvE,nguyen2017novelConvKB,jiang2019adaptiveConvR,vashishth2020interacte} attempt to enhance the interaction between entity and relation embedding vectors in different dimensions. These methods achieved good results on many datasets, but they still suffer from insufficient feature interaction and interpretability. Therefore, this paper uses sparse layers with adjustable sparsity to replace all dense layers. Sparse layers can be seen as an upgrade to convolutional layers, while at the same time alleviating the overfitting issues faced by dense layers through unstructured pruning ~\cite{dettmers2019sparse}.

 In addition, the research of ComDensE ~\cite{kim2022comdense} shows that the effect of a single wide network layer is even better than  a deep network,
which has the degradation problem.
 This paper introduced residual connections ~\cite{he2016deep} to solve the degradation problem when deepening network models. In summary, the contributions of this paper are listed as follows:
 \begin{itemize}
   \item We propose a novel link prediction model for knowledge graph completion, which introduces a \textit{shallow but wide} dynamic layer and a relation-aware layer to the encoding end and a \textit{deep but thin} residual structure to the decoding end. This enables neural networks to perform better information fusion and has the potential to deepen the network  layers.
   \item By replacing all the fully connected layer with sparse layers, our model not only mitigates overfitting risks effectively, but also preserves its capability of feature interactions. Moreover, at comparable interaction levels, fixed sparse structures demonstrate enhanced predictability compared to other methods like dropout or downscale of output dimensions.
   \item A serial of tests and ablation studies were conducted on FB15k-237, WN18RR, and YAGO3-10 demonstrate that our proposed model achieves state-of-the-art performance in terms of Hits@1. Furthermore, by applying t-SNE dimensionality reduction to the output of the gating layer within the dynamic layer, it was discovered that the gating structure distributes weights to expert blocks based on the semantic information of entity-relation pairs.
   % A series of ablation studies and comparative experiments further elucidate the interconnections and efficacy of different components within the model.
 \end{itemize}

 \section{Background and Related Works}

A knowledge graph is a collection of triples (facts) that represent relationships between entities, denoted as $\mathcal{G} = \{(s, r, o)\} \subseteq \mathcal{E} \times \mathcal{R} \times \mathcal{E}$, where $s \in \mathcal{E}$ and $o \in \mathcal{E}$ are the triple subject and object, respectively, and $r \in \mathcal{R}$ is the relationship between them.
Link prediction for KG completion can be viewed as a point-wise learning to rank problem, where the goal is to learn a scoring function that maps an input triple $ (s, r, o)$ to a score $\psi(\cdot)$:
$\mathcal{E} \times \mathcal{R} \times \mathcal{E} \mapsto \mathbf{R}$.

% $\mathcal{E} \times \mathcal{R} \times \mathcal{E} \mapsto$.
%Typical scoring functions are shown in Appendix \ref{Appendix: scoring functions} for the paper length limitation.
Related works on link prediction for knowledge graph completion are summaried as follows:

 \begin{figure*}[t]
   \centering
   \includegraphics[width=1.0\textwidth]{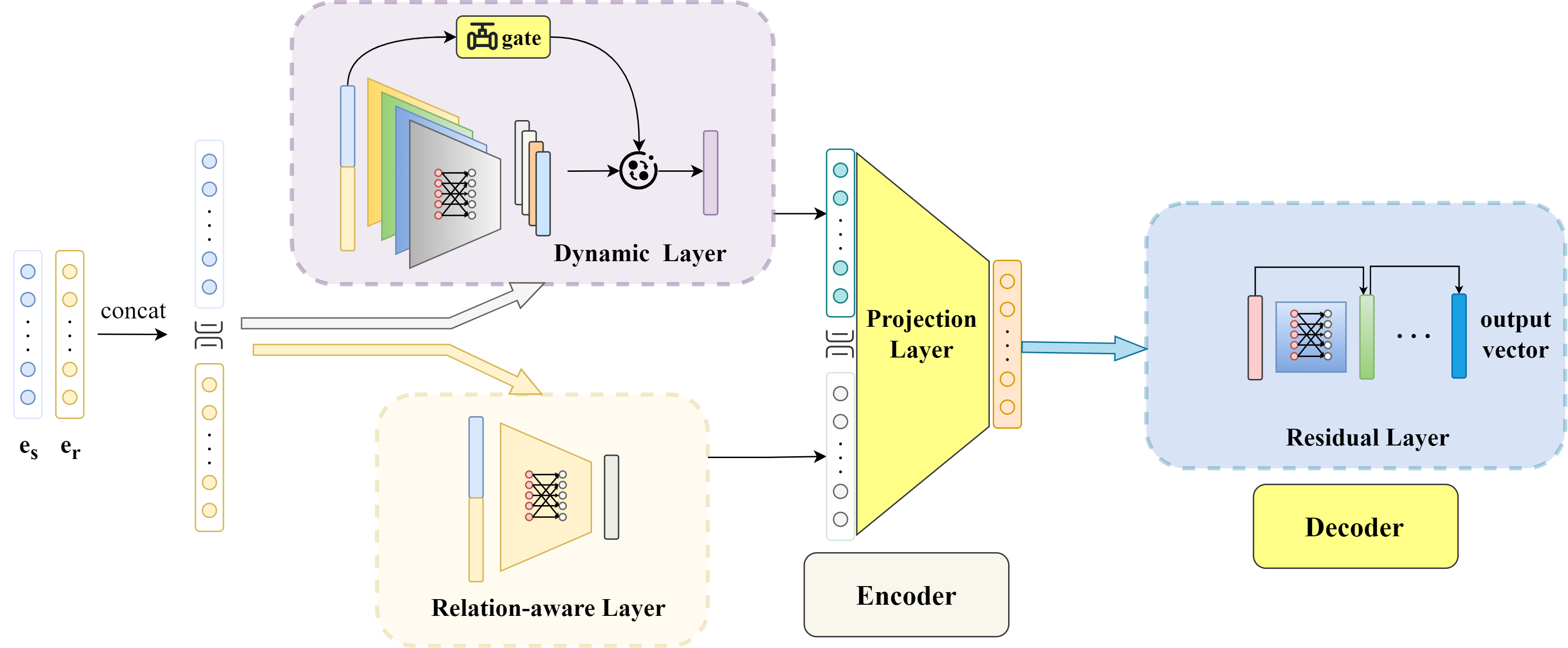}
   \caption{\textbf{The architecture of DSparsE.} The encoding end is composed of a dynamic layer and a relation-aware layer. The decoding end is composed of several residual MLP layers. \textbf{Note that all the dense layers are replaced by sparse layers with certain sparsity degrees.}
    }
   \label{overall architecture}
 \end{figure*}

\label{related works}
\begin{itemize}

   \item \textbf{Tensor decomposition models}

   Tensor decomposition models treat the link prediction as a task of tensor decomposition. It encodes the knowledge graph as a three-dimensional tensor, which is incomplete due to the incompleteness of knowledge graph. This tensor is decomposed into a combination of low-dimensional vectors, from which the embeddings of entities and relations can be obtained.
The model learns the relationships between vectors by setting a loss function and predicts the existence and correlation of underlying facts in  knowledge graph.
Typical tensor decomposition models include DistMult ~\cite{distmult}, ComplEx ~\cite{trouillon2016complex}, TuckER ~\cite{balavzevic2019tucker}, etc.
   Although these models are mostly lightweight and easy to train, they are sensitive to sparse data and have limited modeling capabilities.

   \item \textbf{Translational models}

   Translational models are based on the assumption that the relationship between entities can be represented by the translation of an entity vector. 
A typical translational model is TransE ~\cite{bordes2013translating}. This model learns the embeddings of entities and relations by minimizing the energy function, and predicts the existence of underlying facts in  knowledge graph. Translational models are simple and easy to train, but they are not suitable for modeling symmetric relations and complex relations.
   To address these issues,  TransH ~\cite{wang2014knowledgetransH}, TransR ~\cite{lin2015learningtransR}, and TransD ~\cite{ji2015knowledgetransD} models are proposed to enhance the modeling capability by dynamically mapping entities and relations and suppress the homogenization tendency of embedding vectors.
Moreover, some improved methods based on TransE introduce additional computation overhead.

   \item \textbf{Deep learning models}

   Deep learning models for link prediction can be  categorized as
      models based on simple MLPs,
       models based on convolutional neural networks (CNNs),
       models based on graph neural networks (GNNs),
      and  models based on recurrent neural networks (RNNs) or transformers.
In link prediction, these networks usually take the entity and relation embeddings as input, and obtain a vector after encoding and decoding the input data through several neural network linear and nonlinear layers. Networks based on simple MLPs, such as ComDensE ~\cite{kim2022comdense}, add a relation-aware component, which generates different weight matrices for different relations that appear in the training set based on the common layer.
   Models based on CNNs include ConvE ~\cite{dettmers2018convolutionalConvE}, ConvKB ~\cite{nguyen2017novelConvKB}, ConvR~\cite{jiang2019adaptiveConvR}, and InteractE ~\cite{vashishth2020interacte}. These methods convert embedding vectors into two-dimensional feature maps in different ways and apply filters for convolution. More specifically, ConvR uses the relation embedding vector as the convolution kernel, while InteractE enhances the interaction between features by reshaping them into a checkerboard pattern.
   Models based on GNNs include R-GCN ~\cite{schlichtkrull2018modeling} and CompGCN ~\cite{vashishth2019composition}. These methods use graph convolutional networks to grab the neighborhood information of entities and relations and aggregate them into the entity embedding vector. Those methods natually take advantage of the graph structure and achieve good results on some datasets. However, the parallelization challenge caused by the heterogeneous graph structure limits the performance of these methods.
   Some methods  use fine-tuning pre-trained language models for link prediction, such as KG-BERT ~\cite{yao2019kg} and Rhelphormer ~\cite{bi2022relphormer}.
Although these models achieve good performance, they suffer from high complexity and require external information beyond  knowledge graph.

 \end{itemize}

 \section{DSparsE for Link Prediction}

 This paper proposes a novel neural network model called DSparsE for link prediction,
 whose architecture   is shown in Figure \ref{overall architecture}.
 The proposed DSparsE model consists of two parts: encoder and decoder. 
 The encoding part includes
 a dynamic layer, a relation-aware layer and a projection layer. 
 The decoding part is a residual layer.
 Note that the dense  MLPs are replaced by sparse MLP layers in DSparsE, where a sparsity degree is used to measure the sparsity of weight matrix. 
 The sparsity degree is a hyperparameter that can be adjusted according to  dataset.

 In DSparsE, the robustness of the network is enhanced through dynamic module augmentation, which leverages randomly initialized sparse unstructured pruning via a weight matrix combined with joint learning from stacked expert blocks.
 This structure maximizes the network's expressive power and allows the network to deepen to improve its performance. 
 Since deepening the network may lead to a decrease in accuracy, the  residual connections in DSparsE can alleviate this effect.
 
 In Figure \ref{overall architecture}, DSparsE takes the $d$-dimensional head node embedding   $\bm{e}_s$ and the $d$-dimensional relation embedding  $\bm{e}_r$ as inputs. These two embeddings are concatenated to form a $2d$-dimensional vector,
 which is further passed through a dynamic layer and a relation-aware  layer in parallel. 
 The output features of these two layers  are concatenated and further passed through a projection layer. 
 A detailed introduction of the aforementioned layers in DSparsE is as follows.

 \subsection{Dynamic layer}
 \label{dynamic MLP layer}
 The dynamic layer consists of multiple sparse MLP layers and a gate layer.  
 This network structure enhances the robustness of the model and improves the prediction performance. 
 The dynamic layer takes the concatenation of input vectors $[\bm{e}_s;\bm{e}_r]$ and produces $k$ different output vectors $\bm{e}_{out_1}, \bm{e}_{out_2}, ..., \bm{e}_{out_k}$ through $k$ parallel MLP layers. 
 The output vector of the dynamic layer is obtained by taking a weighted combination of these output vectors $\bm{e}_{out_1}, \bm{e}_{out_2}, ..., \bm{e}_{out_k}$. 
 The combination weights are determined by a gate layer, which includes a dense fully connected layer and a softmax layer controlled  by a temperature parameter $t$.
 The output of the gate layer can be denoted by
 \begin{equation}
       \begin{aligned}
          \bm{g} &= \operatorname{softmax}(\Omega_{gate}([\bm{e}_s;\bm{e}_r]/t)). \\
       \end{aligned}
       \label{eq:gate}
 \end{equation}
 where $\Omega_{gate}(\cdot)$ is a affine function.
 Thus the output of dynamic layer can be written as
 \begin{equation}
     \begin{aligned}
         \bm{e}_{out,D} &= \sum\nolimits_{i=1}^{k} g_i \times\bm{e}_{out_i}, \\
     \end{aligned}
     \label{eq:dynamic MLP layer}
 \end{equation}
 where $\bm{g}\triangleq[g_1,g_2,\cdots,g_k]$.
 
 %\begin{figure}[t]
 %  % 第一幅图片
 %   \centering
 %   %  \vspace*{\fill} % 将图片推到垂直中心
 %    \includegraphics[width=0.9\linewidth]{Dynamic.png}
 %   %  \vspace*{\fill} % 将图片推到垂直中心
 %    \caption{\textbf{Dynamic layer architecture.} The dynamic layer is composed of a series of structurally similar MLP blocks. 
 %The input is processed in parallel through these MLP blocks and then combined with weights. The combination weights are determined by a gating layer.}
 %    \label{fig:dynamic1}
 %\end{figure}
 
 \subsection{Relation-aware layer}
 \label{relation-aware MLP layer}
 To achieve more accurate feature extraction, we introduce a sparse relation-aware layer that changes dynamically with the input relation ~\cite{ji2015knowledgetransD,kim2022comdense}. This can be viewed as part of the network dynamic nature (See~\cite{kim2022comdense} for more details). The output of the sparse realtion-aware layer is given by
 \begin{equation}
       \begin{aligned}
          \bm{e}_{out,R} &=f(\Omega_R^\alpha([\bm{e}_s;\bm{e}_{r}])).
       \end{aligned}
       \label{eq:relation-aware MLP layer}
 \end{equation}
 where $\Omega_R^\alpha(\cdot)$ is a sparse affine function with sparsity degree $\alpha\in(0,1)$, and  $f$ denotes the the activation function.

 \subsection{Projection layer}
 The projection layer in DSparsE is a sparse MLP layer.
 With the input being the vecor concatenation of $\bm{e}_{out,D}$ and $\bm{e}_{out,R}$,
 the output of the projection layer is a $d$-demensional vector and can be given by
 \begin{equation}
       \begin{aligned}
          \bm{e}_{out,P} &=f(\Omega_P^\alpha([\bm{e}_{out,D};\bm{e}_{out,R}])),
       \end{aligned}
       \label{eq:relation-aware MLP layer}
 \end{equation}
 where $\Omega_P^\alpha(\cdot)$ is a sparse affine function with sparsity degree $\alpha\in(0,1)$.

 \subsection{Residual layer}
 \label{residual}
 % \begin{figure}[t]
 %       \centering
 %       \includegraphics[width=0.45\linewidth]{resMLP.png}
 %       \caption{Residual connection structure used in DSparsE.}
 %       \label{fig:residualx}
 % \end{figure}

 % 在同一行中插入两幅图片
 
 %   \hfill % 这里添加一些水平间距
   % 第二幅图片
 % \begin{figure}[htbp]
 %     \centering
 %    %  \vspace*{\fill} % 将图片推到垂直中心
 %     \includegraphics[width=0.85\linewidth]{resMLP.png}
 %    %  \vspace*{\fill} % 将图片推到垂直中心
 %     \caption{\textbf{Residual connection structure.} A residual MLP block consists of a sparse MLP layer, a batchnorm layer, an activation layer (ReLU or others), and a residual connection.}
 %     \label{fig:residualx}
 % \end{figure}

 A residual block consists of a sparse MLP layer, a batchnorm layer, an activation layer (such as ReLU), a dropout layer and a residual connection. 
 The input and output  of the residual block has the same  dimension.
 The decoder of DSparsE is a stack of multiple residual blocks, where the output of the $i$-th   residual block is formulated as
 \begin{equation}
       \begin{aligned}
          \bm{e}_{Res_i} &= f(\operatorname{BN}(\Omega_{Res_i}^\alpha(\bm{e}_{Res_{i-1}})) + \bm{e}_{Res_{i-1}}),
       \end{aligned}
       \label{eq:dynamic MLP layer}
 \end{equation}
 where $\operatorname{BN}(\cdot)$ denotes the batch normalization operation and $\bm{e}_{Res_{i}}$ the output vector of the $i$-th residual block,
 and $\Omega_{Res_i}^\alpha(\cdot)$ is a sparse affine function with sparsity degree $\alpha$.
 Note that $\bm{e}_{Res_{0}}$ is set to the output of the projection layer $\bm{e}_{out,P}$.
 As introduced in ~\cite{he2016deep,ma2022rethinking}, residual connections ease the training of deep networks and prevent the degradation of deep networks. 
 %With similar structure to residual convolutional neural networks, the results in Section \ref*{The effect of residual blocks} show that the sparse residual MLPs have a good performance in link prediction.

 \subsection{Sparse structure of MLP}
 %\label{sparse structure}
 
 % \begin{figure}[t]
 %     \centering
 %     \includegraphics[width=1\linewidth]{sparse.png}
 %     \caption{{Single Sparse Layer.} }
 %     \label{fig:gate}
 % \end{figure}

 It is known that a dense layer may have a large number of useless parameters, which lead to poor model generalization and increase training difficulty ~\cite{dettmers2019sparse}.
 Although 
 a convolutional layer is  a sparse and parameter-sharing linear layer and the number of parameters is much lower than that of a dense (fully connected) layer,
 it leads to insufficient information exchange and difficulty in extracting features. Moreover, convolving the feature embeddings of nodes and relations does not have a physical interpretation.
  To tackle this issue, we can introduce sparse MLP layer in DSparE.
 In the training stage, the elements of weight matrix are initialized randomly with zeros in certain probability, which leads to a sparse MLP layer.
 % This is in contrast to the commonly used Dropout strategy ~\cite{srivastava2014dropout}, which randomly sets neurons to zero during training but is not ideal for this task.
 % Figure \ref{fig:gate} shows the intuitive process of the sparse structure derived from both convolutional and dense layers.
 Given the parameters of a dense MLP layer $\bm{W}$ and a sparsity degree $\alpha$, the parameters of a sparse MLP layer $\bm{W}^{\alpha}$ can be formulated as
 \begin{equation}
       \begin{aligned}
          {W}_{i,j}^{\alpha} = \left\{
          \begin{array}{cl}
             0 & \text{with probability } \alpha, \\
             {W}_{i,j} & \text{with probability } 1-\alpha.
          \end{array}
          \right.
       \end{aligned}
       \label{eq:sparse}
 \end{equation}
 A sparse MLP layer can be viewed from two directions.
 On one hand, it can be viewed as the result of a convolutional layer with enhancing interaction and removing weight sharing.
 On the other hand, it can be viewed as the result of pruning a dense layer.
 %This can be illustrated by Figure \ref{fig: sparse forming} in Appendix.

 %In summary, 
 %
 %
 %The output of the projection layer is a $d$-dimensional vector  $\bm{e}_{encode}$.
 %The whole process of the encoder in DSparsE can be formulated as
 %\begin{equation}
 %    \begin{aligned}
 %       & \bm{e}_{encode}\!=\! f(\Omega_P^{\alpha}([f(\Omega_D^{\alpha}([\bm{e}_s;\bm{e}_r])),f(\Omega_R^{\alpha}([\bm{e}_s;\bm{e}_r]))])),
 %    \end{aligned}
 %    \label{eq:encoder}
 %\end{equation}
 %where $\Omega_\alpha(\cdot)$ denotes an afine function with sparse structure, $\alpha$ denotes the sparsity degree, 
 % $f$ denotes the the activation function, and $T$ deontes the transpose operation.
 %$[\bm{e}_s;\bm{e}_r]$ denote the vector concatenation of $\bm{e}_s$ and $\bm{e}_r$.
 %
 %
 %This $d$-dimensional vector is further passed through a decoding layer consisting of multiple residual connections to obtain the output vector $\bm{e}_{decode} = \Omega_\alpha(\bm{e}_{encode}) $. 
 
 Finally, we calculate the scores and loss function.
 The score is obtained by
 taking the dot product of the output of the residual layer $\bm{e}_{decode}$ and the object entity embedding vector $\bm{e}_o$  and  further applying the sigmoid function, which is formulated as
 \begin{equation}
     \begin{aligned}
         \psi &= \sigma(\bm{e}_{decode}\cdot\bm{e}_o),
     \end{aligned}
     \label{eq:score}
 \end{equation}
 where `$\cdot$' denotes the dot product and $\sigma$ denotes the sigmoid function. 
 For the loss function, we adopt the binary cross entropy loss function 
 \begin{equation}
    \mathcal{L} \!= \!-\frac{1}{N}\sum\nolimits_{i} y_i\log \psi(s, r, o_i)\! +\! (1-y_i)\log (1-\psi(s, r, o_i)),
    \label{eq:loss}
 \end{equation}
 where $N$ is the number of entities, $y_i$ is the label of the $i$-th entity, the label $y_i=1$ of the entity $o_i$ if $(s, r, o_i) \in \mathcal{G}$, and $y_i=0$ otherwise.
 
 \begin{table*}[t]
 \centering
 \small
 \caption{\textbf{A comparison of prediction performance on different datasets.} The best result is in \textbf{bold}, and the second best result is \underline{underlined}.}
 
 \renewcommand{\arraystretch}{1.2}
 \resizebox{1.0\textwidth}{!}{
 \fontsize{7}{8}\selectfont
 \begin{tabular}{c|ccc|ccc|ccc}
 \hline
 \multirow{2}{*}{Model} & \multicolumn{3}{c|}{FB15k-237} & \multicolumn{3}{c|}{WN18RR} & \multicolumn{3}{c}{YAGO3-10} \\
 \cline{2-10}
  & Hits@1 & Hits@10 & MRR & Hits@1 & Hits@10 & MRR & Hits@1 & Hits@10 & MRR \\
 \hline
 TransE ~\cite{bordes2013translating} & 0.199 & 0.471 & 0.290 & 0.422 & 0.512 & 0.465 & -- & -- & --\\
 TransD ~\cite{ji2015knowledgetransD} & 0.148 & 0.461 & 0.253 & -- & 0.508 & -- & -- & -- & --\\
 DistMult ~\cite{distmult} & 0.155 & 0.419 & 0.241 & 0.390 & 0.490 & 0.430 & 0.240 & 0.540 & 0.340\\
 % \hline
 CompGCN ~\cite{vashishth2019composition} & 0.264 & 0.535 & 0.355 & \textbf{0.443} & \textbf{0.546} & \textbf{0.494} & -- & -- & --\\
 R-GCN ~\cite{schlichtkrull2018modeling} & 0.151 & -- & 0.249 & -- & -- & -- & -- & -- & --\\
 % \hline
 ConvE ~\cite{dettmers2018convolutionalConvE} & 0.237 & 0.501 & 0.325 & 0.400 & 0.520 & 0.430 & 0.350 & 0.620 & 0.440 \\
 ConvKB ~\cite{nguyen2017novelConvKB} & -- & 0.517 & \textbf{0.396} & -- & 0.525 & 0.248 & -- & -- & -- \\
 TuckER ~\cite{balavzevic2019tucker} & 0.266 & 0.544 & 0.358 & \textbf{0.443} & 0.526 & 0.470 & -- & -- & -- \\
 ComplEx ~\cite{trouillon2016complex} & 0.158 & 0.428 & 0.247 & 0.410 & 0.510 & 0.440 & 0.260 & 0.550 & 0.360 \\
 RESCAL ~\cite{nickel2011three} & \underline{0.269} & \underline{0.548} & \underline{0.364} & 0.417 & 0.487 & 0.441 & -- & -- & --\\
 RotatE ~\cite{nickel2011three} & 0.241 & 0.533 & 0.338 & 0.417 & 0.552 & 0.462 & 0.402 & 0.670 & 0.495 \\
 
 \hline
 KG-BERT ~\cite{yao2019kg} & -- & 0.420 & -- & -- & 0.524 & -- & -- & -- & -- \\
 
 \hline
 ComDensE ~\cite{kim2022comdense} & 0.265 & 0.536 & 0.356 & \underline{0.440} & 0.538 & 0.473 & -- & -- & -- \\
 InteractE ~\cite{vashishth2020interacte} & 0.263 & 0.535 & 0.354 & 0.430 & 0.528 & 0.463 & \underline{0.462} & \underline{0.687} & \underline{0.541} \\
 
 \hline
 \textbf{DSparsE} (proposed) & \textbf{0.272} & \textbf{0.551} & 0.361 & \textbf{0.443} & \underline{0.539} & \underline{0.474} & \textbf{0.464} & \textbf{0.690} & \textbf{0.544} \\
 \hline
 \end{tabular}
 }
 \label{tab:results}
 \end{table*}

 \section{Experiments and Analysis}
\label{experiments and analysis}

\subsection{Datasets and evaluation settings}
\label{datasets and evaluation settings}
In our experiments, we use 1-N training strategy introduced by \cite{dettmers2018convolutionalConvE} to train DSparsE and evaluate the performance of DSparsE on three typicial datasets: FB15k-237 ~\cite{toutanova2015observed}, WN18RR ~\cite{dettmers2018convolutionalConvE} and YAGO3-10 ~\cite{suchanek2007yago}.
Our evaluation of link prediction is conducted in the filtered setting, where we calculate scores for all other potential triples in the test set that are not present in the training, validation, or test set. To generate these potential triples, we corrupt the subjects for object prediction. We use mean reciprocal rank (MRR) and Hits at N (Hits@N) metrics to evaluate the performance of our model on these datasets.
To ensure robust evaluation, we train and evaluate our models five times and average the performance results.
%The details of datasets and experiment settings are given in Appendix. 

\subsection{Prediction performance}
\label{results}
Table \ref{tab:results} shows the performance of the prposed method compared to existing methods. It can be seen that DSparsE reached the state-of-the-art performance on FB15k-237, WN18RR and YAGO3-10 in terms of Hits@1. On FB15k-237,  it achieves a improvement of 2.6\% and 3.4\% in Hits@1 compared to ComDensE and InteractE, respectively. On WN18RR, the improvement is not  significant compared to CompGCN and TuckER, but it still outperformed those models based on translation and deep learning. On YAGO3-10, DSparsE achieves the state-of-the-art performance on all the metrics, which highlights the effectiveness of the proposed model.
Furthermore, DSparsE performs better than those models based on feature convolution. For instance, on FB15k-237, it achieves a improvement of 14.8\% in Hits@1 compared to ConvE and 6.6\% in Hits@10 compared to ConvKB. KG-BERT, a link prediction model based on BERT pre-trained language model, performs average on small knowledge graph like FB15k-237 and WN18RR, and its accuracy is much lower than DSparsE.
It is observed that DSparsE outperforms KG-BERT with a 24\% and 3\% improvement in Hits@10 on FB15k-237 and WN18RR, respectively. 
%For more detailed results, please refer to  Appendix.

\subsection{Ablation studies \& further experiments}
\label{Ablation studies}
\subsubsection{The effect of sparsity degree}
\label{The effect of sparsity degree}
%figure
% \begin{figure}[t]
%    %这是图文混排的环境，r表示图片靠右，width表示宽度，{7.5cm}表示占用的宽度,调整高度可以用height
%  \centering
% %  \vspace{-11mm}
%  \includegraphics[width=0.5\linewidth]{sparsity_degree1.png}
%     %  \vspace*{\fill} % 将图片推到垂直中心
%      \caption{\textbf{Hits@1} of InteractE, ComDensE, and DSparsE under different sparsity degrees on FB15k-237.}
%      \label{effect of sparsity degree}

% \end{figure}

%\begin{wrapfigure}[16]{r}{6cm}
    %这是图文混排的环境，r表示图片靠右，width表示宽度，{7.5cm}表示占用的宽度,调整高度可以用height

 %\end{wrapfigure}

\begin{figure}[t]
    \centering
    %  \vspace{-11mm}
     \includegraphics[width=0.7\linewidth]{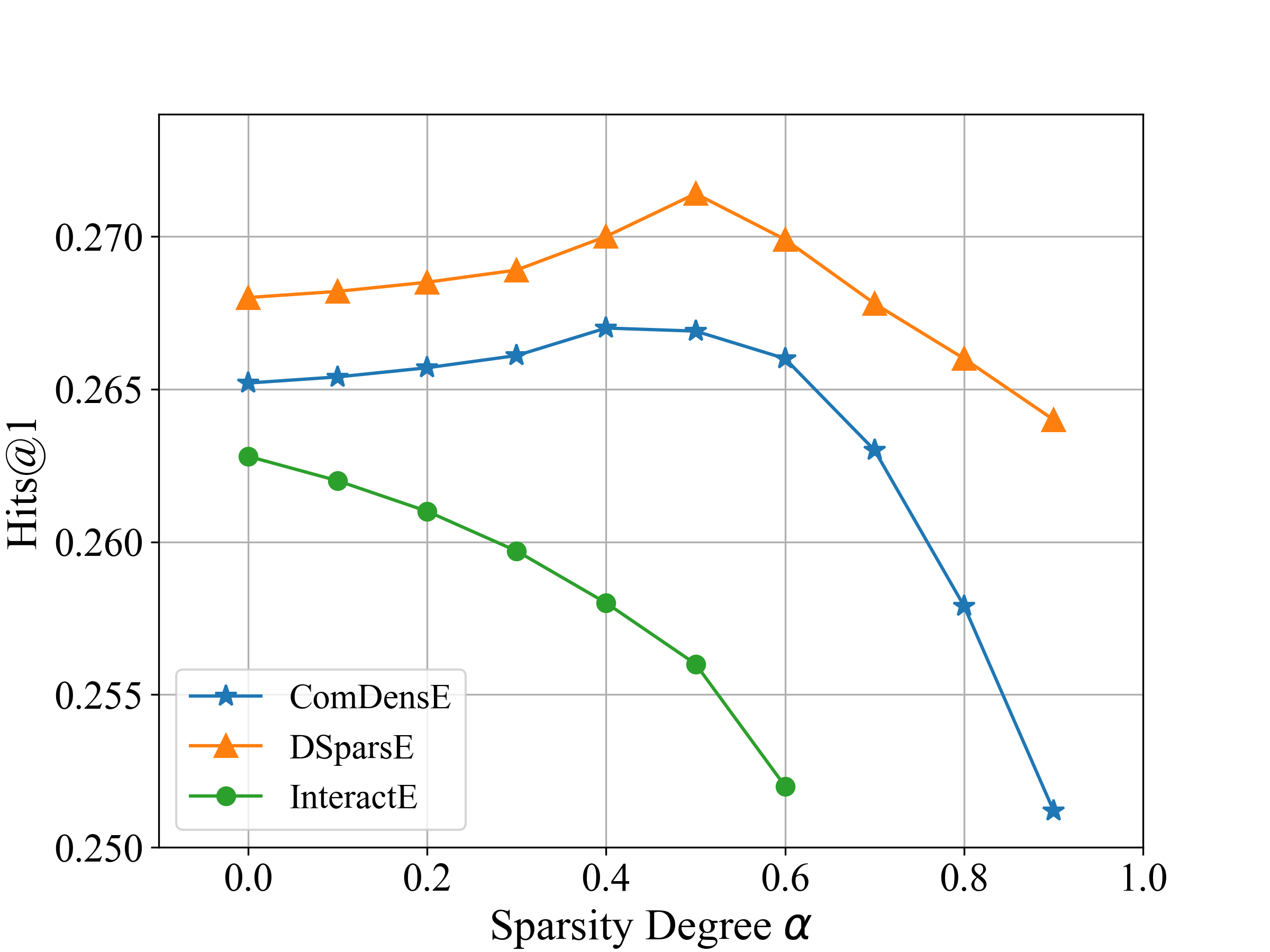}
        %  \vspace*{\fill} % 将图片推到垂直中心
         \caption{\textbf{Hits@1} of InteractE, ComDensE, and DSparsE under different sparsity degrees on FB15k-237.}
         \label{effect of sparsity degree}
 \end{figure}

% The sparsity of neural networks has a significant impact on their performance.
To further explore the effect of sparsity degree, we applied sparse structure to replace the dense layers in DSparsE, ComDensE, and InteractE. The performance comparison is shown in Figure \ref{effect of sparsity degree}.
It can be observed that the accuracies of both ComDensE and DSparsE models first increase and then decrease with the increase of sparsity, and the highest accuracies of both models appear at a sparsity of around 0.5. This indicates that low sparsity in the network can lead to overfitting, limiting its potential, while  increasing sparsity properly can mitigate this issue.
However, excessively high sparsity reduces the number of effective parameters and disrupts neuron connections, diminishing the network's expressive power and impairing training due to decreased neuronal interaction.
Note that DSparsE is less adversely affected by the increase of sparsity compared to ComDensE, owing to its marginally greater parameter count and more complex structure.

On the other hand, the performance of the InteractE model demonstrates a consistent decrease with the increase of sparsity. This trend is due to the model architecture of InteractE, where the final feature decoding layer is only an MLP layer. The experiment results indicate that introducing increased sparsity over the sparse interactions already captured by the earlier convolutional layers adversely affects the model's predictive performance.

Furthermore, the results demonstrate that enhancing a network's effectiveness can be achieved by introducing random sparsity. However, two questions arise:
\begin{enumerate}[leftmargin=*]
   \item  \textbf{Can we achieve a similar performance by reducing the scale of the linear layer?}
   \item  \textbf{Can we achieve a similar performance by increasing the dropout probability?}
\end{enumerate}

\begin{figure*}[htbp]
    \begin{minipage}[htbp]{0.47\linewidth}
    % \begin{figure}[htbp]
        % 第一幅图片
        \centering
%  \vspace{-10mm}
        \includegraphics[width=0.98\linewidth]{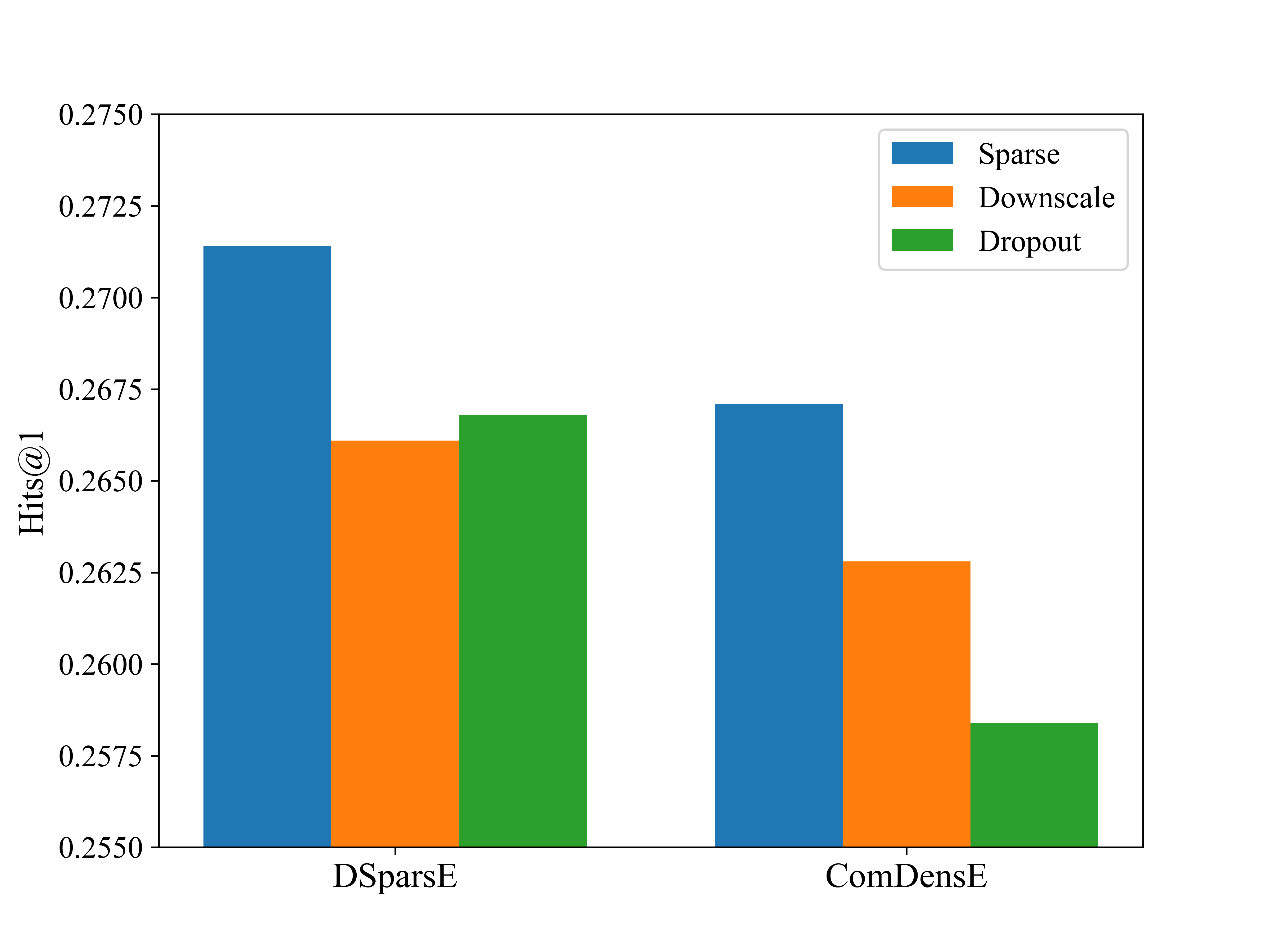}
        \caption{\textbf{The effects of downscale and dropout.} \textit{Sparse} represents the proposed   sparse structure, \textit{Downscale} means cutting off part of the output dimension of the network, and \textit{Dropout} means adding extra dropout based on the original dropout layer. The experiment is conducted on FB15k-237.}
        \label{sparse_research}
    \end{minipage}
    %  \end{figure}
        % \end{minipage}
     %    \end{minipage}
        \hfill % 这里添加一些水平间距
        % 第二幅图片
     %    \begin{minipage}[t]{0.49\linewidth}
    %  \begin{figure}[htbp]
    \begin{minipage}[htbp]{0.49\linewidth}
        \centering
         \vspace{-8mm}
        %  \vspace*{\fill} % 将图片推到垂直中心
         \includegraphics[width=0.98\linewidth]{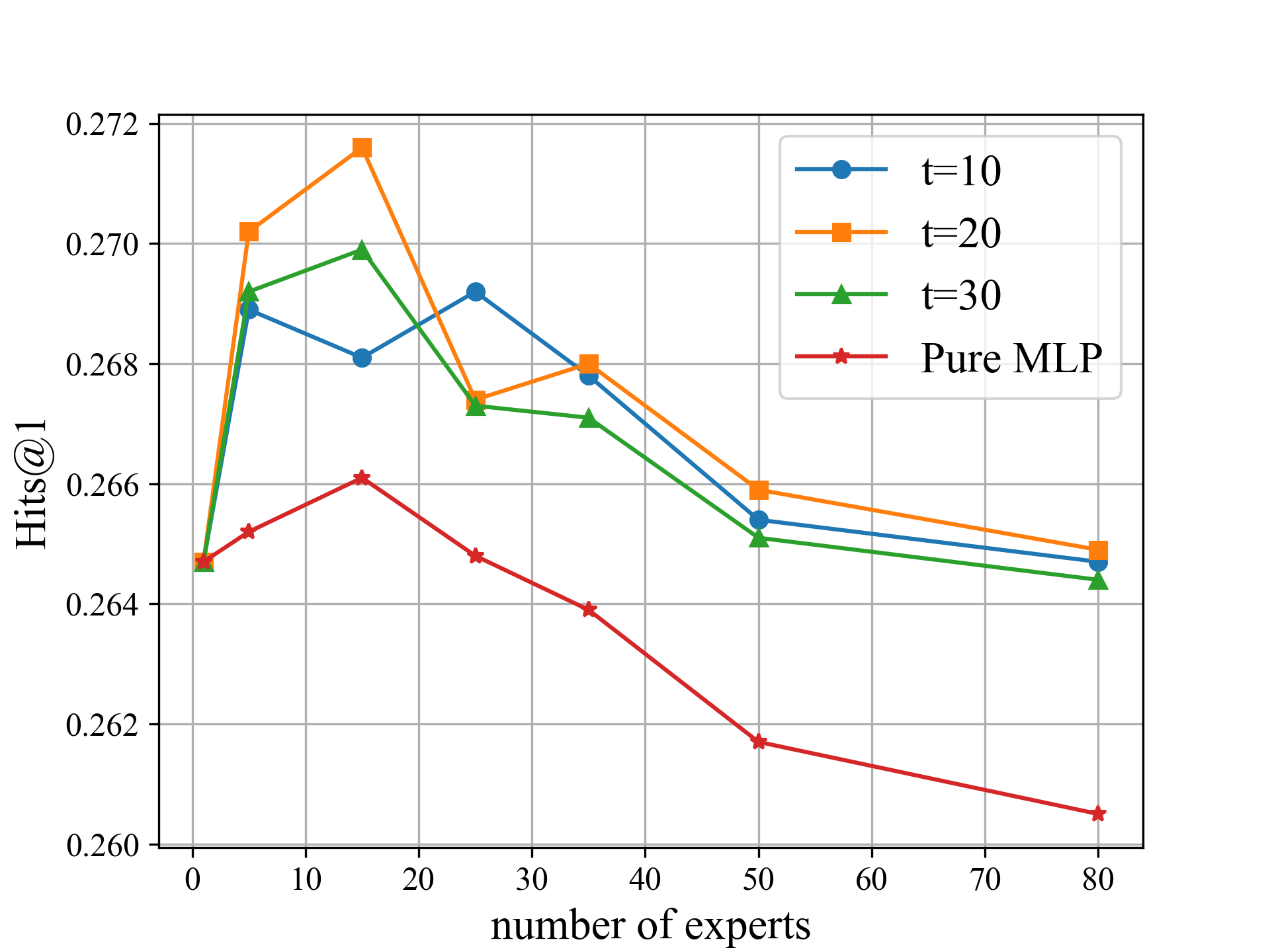}
        %  \vspace*{\fill} % 将图片推到垂直中心
         \caption{\textbf{Hits@1 of DSparsE under different numbers of experts and  temperatures  on FB15k-237.} $t$ denotes the temperature and \textit{Pure MLP} denotes an MLP layer which has the same number of parameters as the dynamic layer.}
         \label{fig:experts}
    \end{minipage}
    \end{figure*}

To address the first question, we do experiments by reducing the output dimension in the linear layers. Specifically, for a linear layer with output dimension $d$, we downscale the output dimension to $\hat{d} = \alpha d$.
For the second question, we do experiments by increasing the dropout rate to $\hat{p} = p + \alpha(1 - p)$, where $p$ is the original dropout rate. The results are shown in Figure \ref{sparse_research}. It indicates that decreasing the number of neurons significantly degrades the performance, whereas  increasing dropout rate drastically deteriorated the performance. This is due to the fact that reducing the neuron number confines the output to a smaller subspace, limiting expressive freedom. On the other hand, since each training iteration changes the dropout mask, an excessively high dropout introduces more uncertainty, thus diminishing network stability.

% \begin{figure}[t]
%    %这是图文混排的环境，r表示图片靠右，width表示宽度，{7.5cm}表示占用的宽度,调整高度可以用height
%  \centering
% %  \vspace{-10mm}
%  \includegraphics[width=0.98\linewidth]{bar.png}
%  \caption{\textbf{The effects of downscale and dropout.} \textit{Sparse} represents the proposed   sparse structure, \textit{Downscale} means cutting off part of the output dimension of the network, and \textit{Dropout} means adding extra dropout based on the original dropout layer. The experiment is conducted on FB15k-237.}
%  \label{sparse_research}
% \end{figure}

\subsubsection{The effect of experts}
\label{The effect of experts}
% \begin{figure}[t]
%     % 第一幅图片
%       \centering
%      %  \vspace*{\fill} % 将图片推到垂直中心
%       \includegraphics[width=0.98\linewidth]{expert_layer.png}
%      %  \vspace*{\fill} % 将图片推到垂直中心
%       \caption{\textbf{Hits@1 of DSparsE under different numbers of experts and  temperatures  on FB15k-237.} $t$ denotes the temperature and \textit{Pure MLP} denotes an MLP layer which has the same number of parameters as the dynamic layer.}
%       \label{fig:experts}
% \end{figure}
% Incorporating expert blocks into the dynamic layer, it enhances dynamics and improves generalization capability of the model.
Figure \ref{fig:experts} illustrates the performance of DSparsE in Hits@1 scores under different  expert and temperature settings  on FB15k-237.
The experiment  results indicate that the prediction performance first increases and then decreases with the rising number of expert blocks.
The  increase of performance  be explained from two aspects. On one hand, in contrast to a non-partitioned fully connected structure (i.e., a very wide fully connected layer), the expert blocks in the dynamic layer represent a form of regular sparse connections.
% (See detailed explaination in Appendix).
These sparse connections are further integrated through a decision layer, namely a gating layer, forming a hypernetwork structure, which brings robustness to the entire network. On the other hand, the expert blocks in the dynamic layer can be viewed as sub-modules in an ensemble learning framework. This ensemble learning architecture can effectively suppress the propagation of errors, reducing the variance in prediction results.

However, when the number of expert blocks becomes large, the performance deteriorates. This is due to an increase in network parameters introduces additional training complexity, diminishing the network's generalization performance. Moreover, the gating network is fundamentally a multi-classifier. An excessive number of categories increases the decision-making complexity of the network.

Another key factor is the temperature of the dynamic layer. High temperature values lead to weight homogenization. Conversely, low temperature values can render many experts ineffective in learning, thus degrade the performance.

% \begin{figure*}[htbp]
%     \begin{minipage}[b]{0.47\linewidth}
%     % \begin{figure}[htbp]
%         % 第一幅图片
%         \centering
%         %   \vspace{3mm}
%            %  \vspace*{\fill} % 将图片推到垂直中心
%            \includegraphics[width=0.98\linewidth]{visualization.png}
%            %  \vspace*{\fill} % 将图片推到垂直中心
%            \caption{\textbf{The output of gated layer for each entity-relation pair.} Each point represents an entity-relation pair in latent space after t-SNE reduction. The color of a point represents the relation type.}
%            \label{relation_pattern}
%     \end{minipage}
%     %  \end{figure}
%         % \end{minipage}
%      %    \end{minipage}
%         \hfill % 这里添加一些水平间距
%         % 第二幅图片
%      %    \begin{minipage}[t]{0.49\linewidth}
%     %  \begin{figure}[htbp]
%     \begin{minipage}[b]{0.47\linewidth}
%          \centering
%          \includegraphics[width=0.91\linewidth]{entity_distribution.png}
%          %  \vspace*{\fill} % 将图片推到垂直中心
%          \caption{\textbf{The distribution of different entities in the same relation cluster (e.g., a relation named \textit{Location}).} The points that close to each other are semantic smilar in latent space.}
%          \label{entity_pattern}
%     \end{minipage}
%     \end{figure*}

\subsubsection{The effects of dynamic layer and relation-aware layer}

Our ablation studies demonstrate that
 both the dynamic and relation-aware layers are essential for achieving the optimal performance, as shown in Table \ref{tab:ablation on dynamic layer and relation-aware layer}. The Dynamic layer compensates for the relation-aware layer's lack of interconnectedness, facilitating the integration of diverse relational knowledge. The expert layer's gating output is determined by head-relation pairs, fostering a more entity-aware weighting system and enabling the connection of different knowledge types. The interaction  between these two layers yields enhanced performance, highlighting their synergistic effect. Furthermore, if both the dynamic layer and relation-aware layer are removed and only the decoder with the residual connection is left, it leads to a significant performance degradation. This decline in performance cannot be mitigated by increasing the number of layers in the decoder. The result indicates that both the encoder and the decoder are indispensable for link prediction in DSparsE.

% \begin{figure}[htbp]
%     \centering
%     %  \vspace*{\fill} % 将图片推到垂直中心
%    %  \vspace{-5.5mm}
%     \includegraphics[width=0.9\linewidth]{bar_expert.png}
%     \caption{\textbf{The effect of Dynamic layer and Relation-aware layer}. The result shows a significant difference when removing the dynamic layer and the relation-aware layer, indicating that both layers are essential to the model. The experiment is conducted on FB15k-237.}
%     \label{expert_ablation}

% \end{figure}

% \captionsetup[table]{font={blue}} % 仅将表的标题设置为蓝色
\begin{table}[t]
   \centering
   \small

   \caption{\textbf{The ablation study on dynamic layer and relation-aware layer on FB15k-237.} \textit{D}, \textit{R} and \textit{Res} denote the dynamic layer, relation-aware layer, and residual layer, respectively.}
   \label{tab:ablation on dynamic layer and relation-aware layer}
   % Increase the padding inside the cells
   \setlength\tabcolsep{6pt} % default value: 6pt
   % Change the font size for the table
   % You can use \small, \footnotesize, etc.
   \begin{tabular}{lcccc}
   \toprule
               & Hits@1    & Hits@10  & MRR \\
   \midrule
   \textbf{D + R + Res (Proposed)}    & \textbf{0.272}    & \textbf{0.551}     & \textbf{0.361}  \\
   \midrule
   D + Res  & 0.254 (-0.018)        & 0.526        & 0.345       \\
   \midrule
   R + Res     & 0.266 (-0.006)    & 0.538     & 0.355 \\
   \midrule
   Res($depth=1$)     & 0.237 (-0.035)    & 0.499     & 0.325   \\
   Res($depth=3$)      & 0.236 (-0.036)    & 0.509     & 0.325   \\
   Res($depth=5$)      & 0.238 (-0.034)    & 0.511     & 0.325   \\
   Res($depth=10$)    & 0.235 (-0.037)    & 0.504     & 0.324   \\
   \bottomrule
   \end{tabular}

\end{table}

\begin{figure}[t]
    % \begin{figure}[htbp]
        % 第一幅图片
        \centering
        %   \vspace{3mm}
           %  \vspace*{\fill} % 将图片推到垂直中心
           \includegraphics[width=0.6\linewidth]{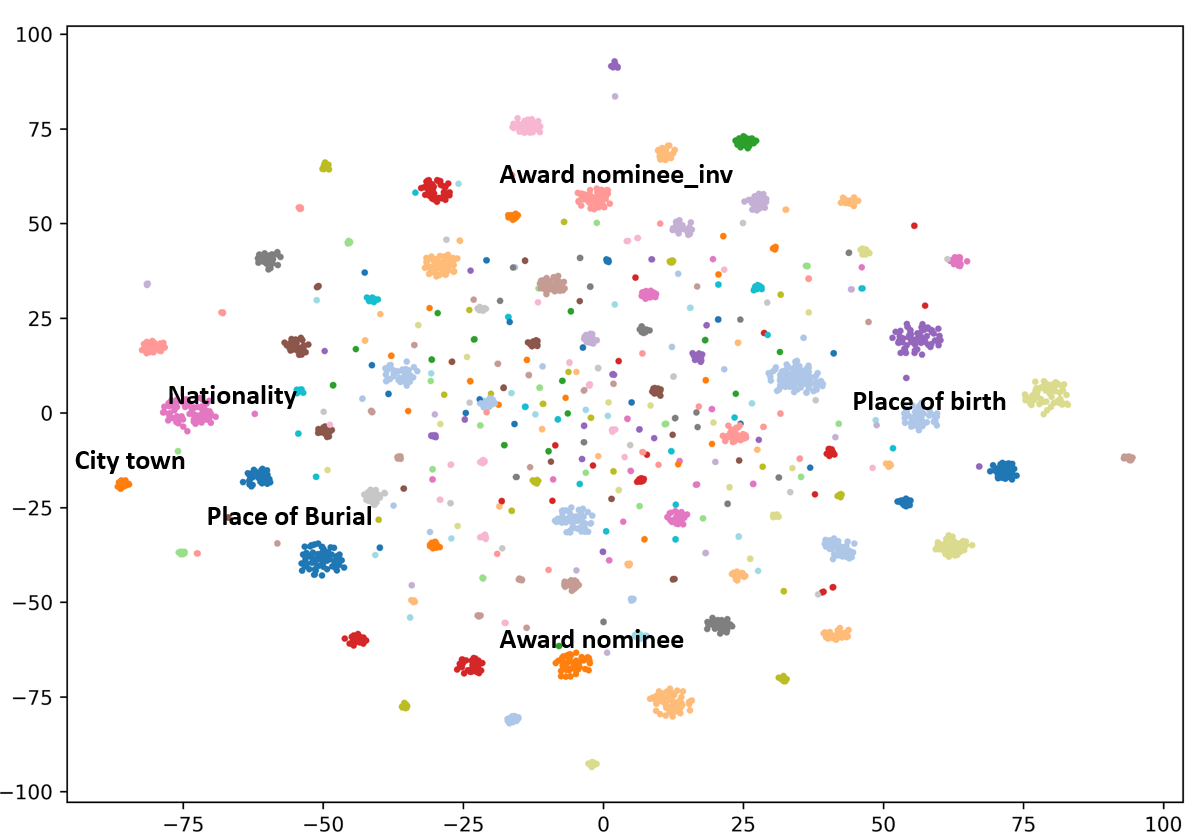}
           %  \vspace*{\fill} % 将图片推到垂直中心
           \caption{\textbf{The output of gated layer for each entity-relation pair.} Each point represents an entity-relation pair in latent space after t-SNE reduction. The color of a point represents the relation type.}
           \label{relation_pattern}
    \end{figure}

    \begin{figure}[h]
        % \begin{figure}[htbp]
            % 第一幅图片
            \centering
            \includegraphics[width=0.6\linewidth]{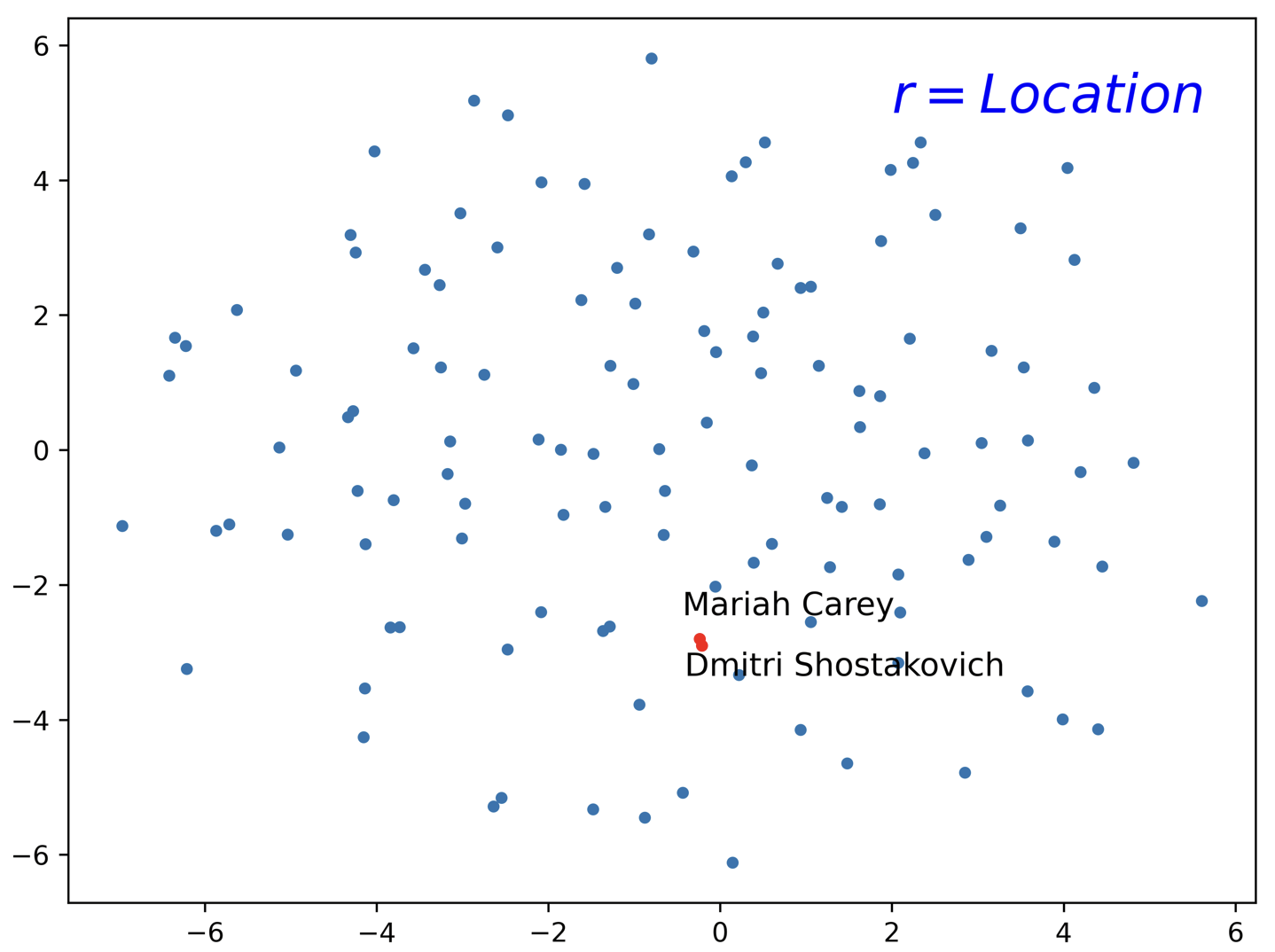}
            %  \vspace*{\fill} % 将图片推到垂直中心
            \caption{\textbf{The distribution of different entities in the same relation cluster (e.g., a relation named \textit{Location}).} The points that close to each other are semantic smilar in latent space.}
            \label{entity_pattern}
        \end{figure}

Furthermore, we disclose some insights of the gate layer. Each entity-relation pair passed through the gating layer yields an output vector. Using the t-SNE dimensionality reduction technique, these high-dimensional vectors can be visualized in Figure \ref{relation_pattern} and Figure \ref{entity_pattern}. Each point in these figures represents a unique entity-relation pair, distinguished by  different colors corresponding to different relationships. The visualization results reveal the following observations:

\begin{itemize}
% \centering
\item The entity-relation pairs with the same relationship type tends to cluster together, which indicates the proximity within the space of the gated layer outputs.

\item The spatial distribution of clusters is significantly influenced by the nature of the relationships.
For instance, relationships denoting inverse meanings (e.g., nominee\_inv and nominee) or semantic opposites (e.g., place of birth vs. place of burial) exhibit divergence in the reduced dimensional space. Conversely, relationships with similar semantics (e.g., nationality and city town) are close in the latent space. This demonstrates that DSparsE can capture various associations between entities and relations.
\item Alterations in the head entity of a relation pair result in minor shifts within the vector output.
For a fixed relation, the relative positions of entity within its corresponding cluster does not display a discernible pattern. This is due to the relatively lower frequency of triples involving individual nodes compared to those associated with a particular relation type, posing challenges in accurately modeling semantic information ~\cite{bordes2013translating}. However, certain examples, such as Mariah Carley and Dmitri Shostakovich (notable in the music domain) demonstrate proximity within clusters pertaining to specific relations.
\end{itemize}

% The result deteriorates dramatically with the depth increases if no residual structure are used. The same thing happens if the deep structure is replaced with a wide linear layer but not as severe as without residual connections.

\subsubsection{The effect of residual blocks}
\label{The effect of residual blocks}

\begin{table}[t]
   \centering
   \small
   % \captionsetup{labelfont={blue}}
   \caption{\textbf{Hits@1 of DSparsE under different residual layer depths on FB15k-237.} Note that the \textit{Wide linear layer} means a wide and shallow network with the same number of parameters
 replaces the deep structure with residual connections in decoder.}
   \label{tab:residual}
   \renewcommand{\arraystretch}{1.2}
   \setlength\tabcolsep{2pt}
   \begin{tabular}{lccc}

   \toprule
   Depth of & With residual  & Without residual  & Wide linear \\
   layers & connection & connection & layer \\
   \hline
   $depth=1$ & 0.2682 & 0.2671 (-0.0011) & --\\
   $depth=2$ & 0.2691 & 0.2522 (-0.0169) & 0.2633 (- 0.0058)\\
   $depth=3$ & \textbf{0.2716} & 0.2276 (- 0.0440) & 0.2550 (-0.0166)\\
   $depth=4$ & 0.2681 & 0.1908 (-0.0773) & 0.2511 (-0.017)\\
   \midrule
   $depth=100$ & 0.2639 & 0.02338 (-0.2401) & 0.2490 (-0.0149)\\
   \bottomrule
   \end{tabular}

   \label{tab:my_label}
\end{table}
% To address the challenge of deepening the network and fully utilize the feature information extracted and fused by the encoder, residual connections are introduced in the decoding end as a final layer before the output. The presence of residual connections allows for further deepening of the network without compromising performance, thereby ensuring enhanced expressive capability.

With the increasing scale of dataset, a deeper decoding layer is expected to ensure that the network's performance. However, simply increasing MLP layers leads to a rapid degradation in performance.
Employing residual connections maintains the expressive potential of the network effectively.
The presence of residual connections in DSparsE ensures the expressive capability when deepening the network. Table \ref{tab:residual} shows the performance of DSparsE in Hits@1 under different numbers of residual blocks  on FB15k-237.
It can be observed that the accuracy decreases rapidly as the number of layers increases if the residual blocks are replaced with fully connected layers. 
In our model, the residual connections reduce the effect of increasing the number of layers.
Furthermore, if a wide and shallow network with the same number of parameters is used to replace the deep structure with residual connections, there is still a performance degradation.
These experiments demonstrate that a \textit{shallow but wide} encoder for feature extraction and a \textit{deep but thin} decoder for feature decoding can effectively enhance the performance of link prediction.

\section{Conclusion}
\label{conclusion}
This paper proposed a new model called  DSparsE  for knowledge graph completion.
By introducing wide dynamic layer and relation-aware layer as an encoder and a deep residual connection layer as a decoder, the model representation power was  enhanced effectively.
This model employs sparse MLP layers and  residual structures to alleviate overfitting, which reduces the difficulty of training deep networks and improves prediction performance.
The experiment results demonstrate that the proposed DSparsE achieves the best performance in terms of Hits@1 on the FB15k-237, WN18RR, and YAGO3-10 datasets.
Moreover, it was discovered that the hypernetwork structure formed by gated layers can effectively capture the semantic features and semantic associations of entity-relation pairs, with the results of latent space dimensionality reduction exhibiting interesting clustering and intra-cluster deviation phenomena.
Ablation studies have further proven that the \textit{shallow-to-deep} network structure of DSparsE improves  the link prediction performance.

\bibliographystyle{splncs04}
\bibliography{icpr24.bib}

\end{document}